\documentclass[prb,twocolumn,showpacs,aps,floats,floatfix,superscriptaddress]{revtex4}
\usepackage{epsfig}

\begin{document}

\title{Influence of salt and viral protein charge distribution on encapsidation of single-stranded viral RNA molecules}
\author{Antonio \v{S}iber} 
\affiliation{Department of Theoretical Physics, Jo\v{z}ef Stefan Institute, SI-1000 Ljubljana, Slovenia}
\affiliation{Institute of Physics, P.O. Box 304, 10001 Zagreb, Croatia}
\author{Rudolf Podgornik}
\affiliation{Department of Theoretical Physics, Jo\v{z}ef Stefan Institute, SI-1000 Ljubljana, Slovenia}
\affiliation{Department of Physics, University of Ljubljana, SI-1000 Ljubljana, Slovenia}
%
%

\begin{abstract}
We examine the limits on viral composition that are set by the electrostatic interactions effected by 
the charge on the viral proteins, the single-stranded viral RNA molecule and monovalent salt ions in the solution. 
Within the mean-field model of viral energetics we demonstrate the prime importance 
of the salt concentration for the assembly of a virus. 
We find that the encapsidation of the viral RNA molecule is thermodynamically suppressed in solutions 
with high concentrations of monovalent salt. This effect is significantly less important in viruses with 
proteins whose charge distribution protrudes into the interior of the capsid, leading to an increase 
in the stability of such viruses in solutions with high salt concentrations. The delocalization 
of positive charge on the capsid protein arms thus profoundly increases reliability of 
viral assembly in high-salt solutions. 
\end{abstract}
\maketitle

Viruses are a prime example of precise spontaneous assembly. It was more than fifty years ago since Fraenkel-Conrat and 
Williams demonstrated that fully infectious tobacco mosaic viruses (TMV) could be created simply 
by mixing the viral RNA molecules together with the viral 
proteins\cite{Fraenkel}. Under the right conditions (pH and salinity), the viruses formed spontaneously, i.e. 
without any special external constraints. 

There is no unanimous view concerning the physical interactions that guide the viral self-assembly. 
It is often supposed that {\em specific} interactions acting between the viral genome and its proteins 
guarantee the precise assembly. However, this cannot be the entire story since it has been 
demonstrated that {\em (i)} empty viral protein coatings (capsids) assemble, at least when the amount of salt in the bathing solution is large 
enough \cite{Iwasaki,Ganser}, and {\em (ii)} the filled virus-like  particles form even when the viral genome is replaced by noncognate RNA 
molecules \cite{Bancroft,Ganser}. 

The role of salt in the viral assembly is of essential importance. 
Already in the early studies of TMV assembly \cite{KlugPRS} it was found that the viral proteins 
can be assembled in capsid-like structures 
at sufficiently high salt concentrations even if the pH of the solution is high enough to prohibit the assembly in 
low ionic concentrations (or causes an alkaline degradation of the assembled viruses). It is experimentally 
documented \cite{cuillel1} that upon lowering the pH of the high-salt solution ($>$ 0.8 M LiCl) with 
disassembled proteins and RNA of brome mosaic virus below pH 6.5, only {\em empty} viral capsids form. 
All these experimental findings clearly indicate that there is a nonspecific interaction of  
electrostatic origin acting between the viral proteins and RNA molecules. This interaction depends crucially on the concentration of salt ions in the bathing solution. 

The aim of this letter is to decipher the role of salt in the assembly of viruses that contain single-stranded 
highly negatively charged (one elementary charge per nucleotide) RNA (ssRNA) molecule. The 
proteins of such viruses typically carry positive net charge \cite{MuthuPNAS} at physiological pH. For many of these 
relatively simple viruses it has been experimentally demonstrated that they can be spontaneously assembled {\em in vitro}. 
Spontaneous assembly takes place only if the energy of the capsid/genome complex is 
favorable - as proposed already by Caspar and Klug \cite{casparklug}. In the 
context of electrostatic interactions, this means that the number of charges on the ssRNA and 
the capsid must be related and this relation should depend also on the amount of added salt. 

We approach the problem by representing the viral ssRNA as a generic flexible polyelectrolyte 
with effective monomer size $a$, and $pe$ charge per monomer, where $e$ is the electron charge and $0 < p < 1$.  The polyelectrolyte concentration, $\Psi(r)^2$ 
and electrostatic potential, $\Phi(r)$, are treated as continuous real-valued fields that 
minimize the mean-field ground state 
dominance free energy of the polyelectrolyte/capsid/salt system\cite{Andelman1}, $F$, 
in the subspace of fixed total number of polyelectrolyte monomers, $N$, so that  
\begin{equation}
F=\int f(r) d^3r - \mu \left ( \int d^3 r \Psi ({\bf r})^2 - N \right ), 
\end{equation}
where $\mu$ is the Lagrange multiplier enforcing the condition of fixed number of monomers, and 
\begin{widetext}
\begin{eqnarray}
f(r) &=& k_B T \left[ \frac{a^2}{6} (\nabla \Psi(r))^2 + \frac{v}{2}\Psi(r)^4 \right ]
+ \left[ e c^{+}(r) - e c^{-}(r) - p e \Psi(r)^2 + \rho_p (r) \right] \Phi(r) 
- \frac{\epsilon_0 \epsilon}{2} (\nabla \Phi(r))^2 \nonumber \\
&+& \sum_{i= \pm} \left \{ 
k_B T \left [ 
c^{i}(r) \ln c^{i}(r) - c^{i}(r) - \left( 
c_0^{i} \ln c_{0}^{i} - c_{0}^{i}
\right)
\right ] \right.
- \left .\mu^{i} \left[ 
c^{i}(r) - c_{0}^{i}
\right]
\right \}.
\label{eq:fions}
\end{eqnarray}
\end{widetext}
Here $T$ is the temperature, $k_B$ is the Boltzmann constant, 
$c^{\pm}$ are the concentrations of $+$ and $-$ monovalent salt ions, 
with $c_0^{\pm}$ being their 
bulk concentrations, and $\mu^{\pm}$ their chemical potentials, $\epsilon \epsilon_0$ is 
the permittivity of water, and 
$v$ is the (non-electrostatic) excluded volume of the polyelectrolyte chain. The 
density of charge located on the capsid  proteins is denoted by 
$\rho_p(r)$. We shall consider this charge density to be fixed, i.e. we shall investigate 
the stability of the {\em assembled} capsids at fixed (physiological) pH. The variation 
of the free energy functional with respect to fields $\Psi$, $\Phi$ and 
$c^{\pm}$ yields two coupled non-linear equations - the generalized polyelectrolyte Poisson-Boltzmann 
equation \cite{rudireview}, and 
the Edwards equation\cite{Edwards1}.
These equations are solved numerically with the requirement that the polyelectolyte density amplitude 
field vanishes at the interior capsid radius, $R$. 
It is known that the asphericity of ''spherical'' (icosahedral) viruses 
increases with the mean radius of the virus 
\cite{Nelson1}. However, deviations from the perfectly spherical shape are generally 
small even for quite large viruses \cite{Nelson1}, thus our approximation of spherical symmetry 
is not expected to be a serious limitation. We do not 
explicitly account for the mechanical elasticity of the polyelectrolyte, {\sl i.e.} our approach 
is taylored  to flexible polyelectrolyte molecules, the ssRNA in particular. Our 
approach cannot account for the details of the RNA conformation, 
such as its branched and locally double stranded structure, or its possible dodecahedral 
ordering in the vicinity the capsid, a feature that 
has been investigated recently \cite{svedi,bruinpress}. 

\begin{figure}[ht]
\centerline{
\epsfig {file=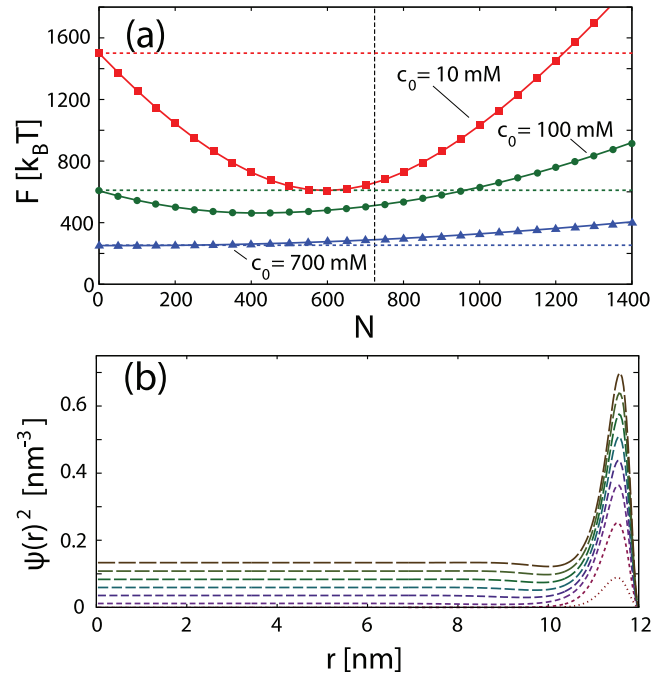,width=8.5cm}
}
\caption{Panel a): The free energies of the polyelectrolyte/capsid/salt system as functions of the number of 
monomers. The parameters used are $p=1$, $a=0.5$ nm, $v=0.05$ nm$^3$, $T=300$ K, 
$\sigma$=0.4 $e$/nm$^2$, and $R$=12 nm. 
The results are shown for three different bulk concentrations of salt ions, 
$c_0=10, 100$, and 700 mM. The vertical dashed line denotes the number of monomers for which the 
polyelectrolyte charge equals in magnitude the capsid charge ($N \approx 724$). Panel b): The polyelectrolyte 
concentration profile for $c_0$=100 mM. The curves displayed correspond to $N$=100, 300, 500, 700, 900, 
1100, 1300, and 1500. The lines are styled so that the length of their dashes is proportional to $N$.
}
\label{fig:fig1}
\end{figure}

Since the size of empty capsids is usually the same as in the fully functional viruses (at least in 
a range of pH and salinity values \cite{Iwasaki,Ganser}), it is reasonable to fix the capsid radius 
at a prescribed value, corresponding to the preferred mean curvature of the empty capsid \cite{SiberPodgornik1} and to examine the 
energetics of the filled capsid depending 
on the amount and type of enclosed polyelectrolyte and the concentration of salt ions in the bathing solution.
Figure \ref{fig:fig1}a) displays the free energies of the polyelectrolyte/capsid/salt system as functions 
of the number of monomers ($N$) in the polyelectrolyte for three different salt concentrations in 
the bulk bathing solution. In this calculation, we have represented the protein charges as a uniformly charged 
infinitely thin spherical shell of surface charge density $\sigma = 0.4$ $e$/nm$^2$ and inner radius 
$R$=12 nm. This should be representative for typical ssRNA viruses \cite{bruin1,Kegel}.
In Fig \ref{fig:fig1}b) we represent the polyelectrolyte concentration profile for 
several polyelectrolyte lengths and for ''physiological'' salt concentration of $c_0=$100 mM.
There are several important messages that can be read directly off this figure. First of all, 
the energetics of viral capsids is profoundly influenced by the concentration of salt. 
Second, there is a {\em critical number of monomers} that can be thermodynamically packed within the capsid. This 
happens at the point when the total energy of the system becomes larger than the 
energy of the empty capsid ($N=0$); in Fig. \ref{fig:fig1} these are the points at which the full lines 
intersect with the dotted horizontal lines for given bulk concentrations of the salt. For polyelectrolytes 
larger than this critical size, formation of {\em empty} capsids is thermodynamically preferable. While it is 
generally easier to form capsids at elevated salt concentration (this is seen from the smaller values of the 
free energy at high salt, irrespective of the number of monomers), it is more difficult to form 
{\em filled} capsids. For low salt 
concentrations, the critical number of monomers is such that the total polyelectrolyte charge is about 
two times larger in magnitude from the charge on the capsid. However, as the salt concentration increases, 
the critical number of monomers decreases. For $c_0=$ 700 mM, it is only about 100, twelve times 
smaller than the critical number in low salt ($c_0=$ 10 mM). We define the optimal number of monomers as the one that 
minimizes the free energy for a given salt concentration (in agreement with Ref. \onlinecite{bruin1}). 
In low-salt solutions, this happens when the total 
polyelectrolyte charge 
approximately equals the capsid charge, but in elevated salt, the optimal number of monomers 
decreases. In contrast to Ref. \onlinecite{bruin1}, we find that the optimal number of monomers is such 
that the magnitude of charge on the polyelectrolyte is always smaller than the protein charge. 

These findings can be better understood by examining the polyelectrolyte density in the capsid [Fig. \ref{fig:fig1}b)]. 
Due to attraction between the polyelectrolyte and the capsid, there is always a maximum in the polyelectrolyte 
concentration at a distance $\cong a$ from the capsid [distributions similar to that shown in Fig. \ref{fig:fig1}b) have 
been experimentally observed - see e.g. Ref. \onlinecite{MuthuPNAS}]. When the number of monomers is larger than the 
optimal one, the polyelectrolyte density becomes finite throughout the capsid, filling the capsid core, although the 
maximum in the density close to the capsid is still distinguishable 
even for $c_0$ as large as 700 mM (not shown). At high salt, the electrostatic interactions are 
screened and the polyelectrolyte entropy becomes 
important in the total balance of free energy. In this regime, it becomes energetically favorable even for 
quite short (depending on exact amount of salt) polyelectrolyte to 
delocalize over the whole capsid as there is enough salt to efficiently screen the capsid charge even in the absence 
of the polyelectrolyte. We find that sub-optimal polyelectrolyte conformations are always such that the polyelectrolyte is located only 
within a shell close to the capsid, while the super-optimal conformations are extended throughout the whole capsid, 
irrespectively of the salt concentration. A similar result was found by authors of Ref. \onlinecite{svedi}, who 
used a discretized version of a model akin to ours that does not include the effects of salt, however.

\begin{figure}[ht]
\centerline{
\epsfig {file=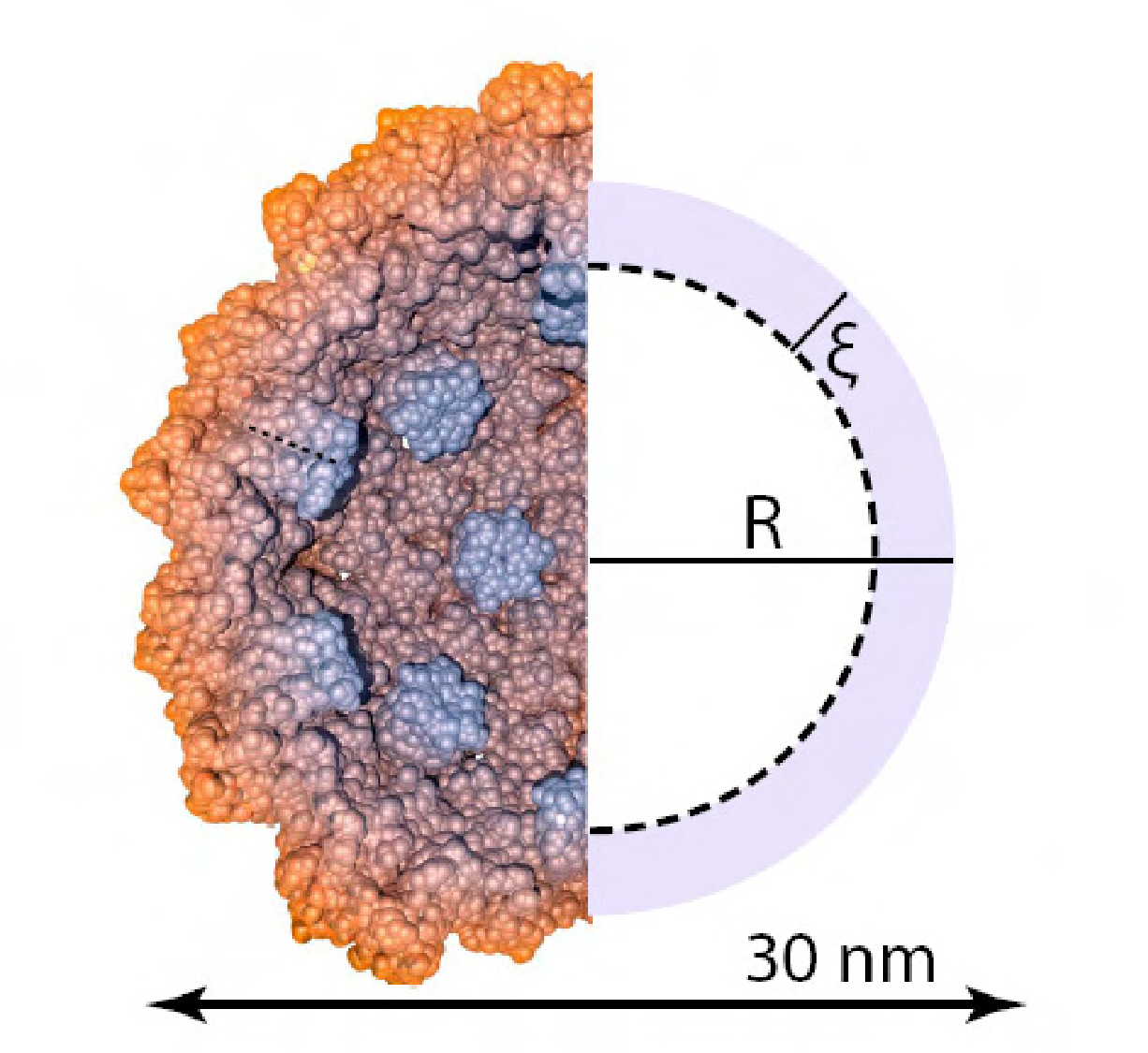,width=8.5cm}
}
\caption{One quarter of the cucumber mosaic virus capsid (strain FNY). The image 
was constructed by applying the group of icosahedral transformations to the RCSB Protein Databank 
entry 1F15 and all atoms in the resulting structure were represented as spheres of radius 3.4 \AA \hspace{0.8mm} (which is the
experimental resolution \cite{cucexp}). They were colored in accordance with their distance from the geometrical center of the capsid.
}
\label{fig:fig2}
\end{figure}

\begin{figure}[ht]
\centerline{
\epsfig {file=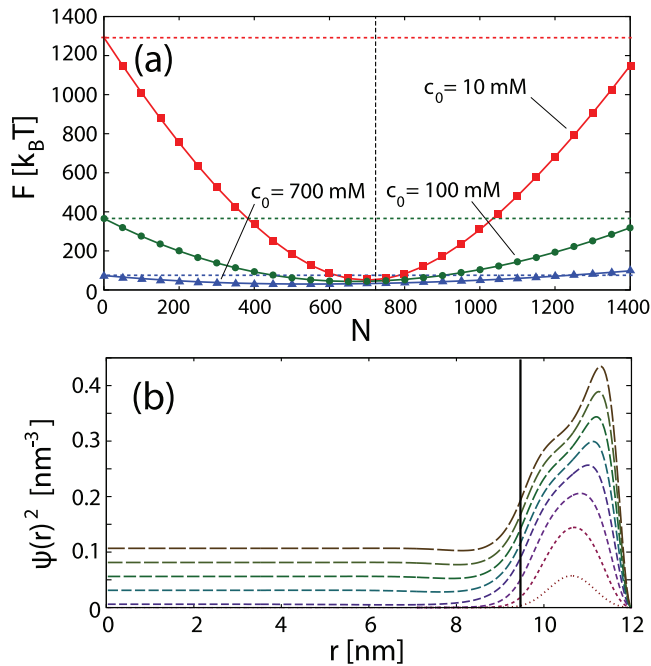,width=8.5cm}
}
\caption{The same as in Fig. \ref{fig:fig1} except for the protein charge density that is represented 
as in Eq. (\ref{eq:stretched}) with $Q_c$ = 724 $e$ and $\xi$ = 2.5 nm.}
\label{fig:fig3}
\end{figure}

Not all ssRNA viruses can be represented by an infinitely thin shell of positive protein charge density. While this may be 
a reasonable approximation for viruses as dengue or yellow fever, it is certainly a poor approximation for e.g. 
cucumber mosaic virus (see Fig. \ref{fig:fig2}), tomato aspermy virus and the much investigated cowpea 
chlorotic mottle virus \cite{cucexp}. These viruses are known to have specifically shaped 
capsid proteins, so that their N-terminal tails are highly positively charged and stretched. 
When virus of this type is fully assembled, the capsid protein 
''arms'' protrude into its interior. It has recently been 
suggested that the existence of highly basic capsid peptide arms can explain the proportionality 
between the net charge on the capsid proteins and the total length of the ssRNA 
viral genome \cite{MuthuPNAS}. Here we are more interested in investigating whether such delocalization of 
the protein charge influences the dependence of stability of a virus on the salt concentration.
We represent the capsid charge density as 
\begin{equation}
\rho_p (r) = Q_c / (4 \pi r^2 \xi), \; R-\xi < r < R,
\label{eq:stretched}
\end{equation}
and $\rho_p (r) = 0$ otherwise, {\sl i.e.} we treat the capsid peptide arms as strongly stretched polyelectrolytes, but not necessarily 
of brush type, of length $\xi$, carrying in total a charge $Q_c$ per capsid. The actual charge distribution in real viruses depends 
on the amino acid content of the capsid peptide arms, but it is worth mentioning that results practically 
indistinguishable from those shown in Fig. \ref{fig:fig3} are obtained by assuming that $\rho_p = {\tt const.}$ for 
$R-\xi < r < R$ and zero otherwise, so that the total protein charge is still $Q_c$ (the robustness of the 
results is mostly due to the fact that $\xi \ll R$). Note that we do not account for steric 
repulsion acting between the capsid peptide arms and the viral ssRNA {\sl i.e.} for the loss 
of interior capsid volume available to ssRNA resulting from the protrusion of pieces of 
capsid proteins into the capsid interior. Figure \ref{fig:fig3} is analogous to Fig. \ref{fig:fig1} but for capsid charge density represented by Eq. (\ref{eq:stretched}). The total charge on the capsid ($Q_c=724$ $e$) is the same in these two cases studied. The arm length chosen ($\xi = 2.5$ nm) should be representative of cucumber mosaic virus (see Fig. \ref{fig:fig2}). 
Comparing Figs. \ref{fig:fig1} and \ref{fig:fig3}, 
one can conclude that {\em (i)} when the polyelectrolyte length
is optimal, the viruses with charge delocalized on the capsid arms 
are bound significantly stronger - this 
can be seen from smaller values of $F$ achieved at the minimum
when compared with the corresponding values in
Fig. \ref{fig:fig1} (for $c_0$=100 mM, the difference in binding energies
is about 450 $k_B T$ per capsid), and 
{\em (ii)} the optimal length of the polyelectrolyte is {\em significantly less}
influenced by the salt concentration (for $c_{0}$ = 10 mM,
100 mM, and 700 mM, the optimal monomer numbers
are 700, 650, and 550, respectively). Our second result is in 
rough agreement with the 
findings by Belyi and Muthukumar who estimate that the total ssRNA 
charge and the total capsid charge are equal up to a quantity of the 
order of 10 elementary charges for salt concentrations below about 
$c_0=100$ mM \cite{MuthuPNAS}. Our results thus bridge the two \cite{bruin1,MuthuPNAS}
apparently contradictory 
previous attempts to describe the viral energetics and show that 
the spatial distribution of protein charge determines the important 
features of the energetics of viruses with regard to salt concentration. 
Note from Fig. \ref{fig:fig3}b) 
that the thickness of the ssRNA ''shell'' is determined by the length of the protein 
arms, unlike in the case of infinitely thin shell of viral protein charge where it 
is determined by $a$ and $v$ parameters of the polyelectrolyte. For superoptimal 
polyelectrolyte lengths there appears a characteristic ''two-humped'' profile of 
the polyelectrolyte concentration which results from an interplay of the four 
length scales involved in this case - the $a$ and $v$ parameters of the 
polyelectrolyte, the length of the capsid arms, $\xi$ and the Debye-H\"{u}ckel 
screening length that depends on the salt concentration \cite{preparation}.

In summary, our results show that the influence of salt on the energetics of 
viruses is quite important, especially for viruses whose positive charge 
on the capsid interior may be well represented as an infinitely thin shell. 
However, the delocalization  of positive charge on the capsid protein arms 
profoundly increases the reliability of viral assembly in high-salt solutions. This effect  
also increases resistance of the assembled viruses towards disassembly in the solutions 
containing high concentration of (monovalent) salt. Our results strongly suggest that the 
highly charged delocalized capsid protein arms may offer an evolutionary 
advantage to viruses that have them. This does not conflict with the fact that the viruses 
we examined are very simple ones, since, as Caspar and Klug already noted \cite{casparklug} 
''viruses could not possibly exist before cells [and] the minimal viruses could be 
considered highly evolved forms''. The delocalized capsid peptide arms may also have a role in the 
kinetics of the assembly, possibly speeding it up, and it is in this respect intriguing 
that we clearly see their effects also in the energetics of the assembled viruses. 

This work has been supported by the Agency for Research and Development of Slovenia
under grants P1-0055(C)  and L2-7080, the Ministry of Science, Education, and Sports of Republic of Croatia  through Project No. 035-0352828-2837,  and by the National Foundation for Science, Higher Education, and Technological Development of the Republic of Croatia through Project No. 02.03./25.

\end{document}